\documentclass[12pt]{article}
\usepackage{amsmath}
\usepackage{graphicx}
\usepackage{epsfig}

\def \ha {{\textstyle{1 \over 2}}}\def \vp {\varphi}\def \J {{\cal }}

\newcommand{\foot}{\footnote}
\newcommand{\bi}{\bibitem}
\newcommand{\ci}{\cite}
\newcommand{\rf}{\eqref}
\newcommand{\ov}{\over}
\newcommand{\x}{\times}

\newcommand{\be}{\begin{equation}}
\newcommand{\ee}{\end{equation}}
\newcommand{\ba}{\begin{array}}
\newcommand{\ea}{\end{array}}
\newcommand{\bea}{\begin{eqnarray}}
\newcommand{\eea}{\end{eqnarray}}
\newcommand{\bp}{\begin{pmatrix}} 
\newcommand{\emp}{\end{pmatrix}}
\newcommand{\bit}{\begin{itemize}}
\newcommand{\eit}{\end{itemize}}
\newcommand{\no}{\nonumber}
\newcommand{\la}{\label}

\def \rX {{\rm X}} 
\def \sql {\sqrt{\lambda}}
\def \adss {$AdS_5 \x S^5$ \ }

\def \te {\textstyle}

\makeatletter
\renewcommand\section{\@startsection {section}{1}{\z@}%
                                   {-3.5ex \@plus -1ex \@minus -.2ex}%
                                   {2.3ex \@plus.2ex}%
                                   {\normalfont\large\bfseries}}
\renewcommand\subsection{\@startsection{subsection}{2}{\z@}%
                                   {-3.25ex\@plus -1ex \@minus -.2ex}%
                                   {1.5ex \@plus .2ex}%
                                   {\normalfont\normalsize\bfseries}}
\def\eps{{\epsilon}}
\def\Tr{{\rm Tr}}
\def \foot {\footnote}
\def \bi{\bibitem}

\def \ci{\cite}
\def \N {{\mathcal N}}

\def \const {{\rm const}}

\def\foot{\footnote}

\def \ci {\cite}
\def \foot {\footnote}
\def \bi{\bibitem}

\def \Tr {{\rm Tr}}

\def \l  {\lambda}
\def \const {{\rm const}}

\def \N {{\mathcal N}}

\def \m {\mu}\def \bi{\bibitem}
\def \la {\label}

\def \J {{\cal J}} 
 
 \def \r {\rho}\def \ov {\over}
\def \edo \end{document}

\def \s {\sigma}
 \def \g {\gamma} 

\def \edd {\end{document}}
\def \fo {{\textstyle {1 \ov 4}}}
\def \fot {{\textstyle {3 \ov 4}}}
\def \C {{\cal C}}
\def \sql  {\sqrt{\l}}

\def \rX {{\rm X}} 
\def \lan {\langle}
\def \ran {\rangle}
\def \arctanh {{\rm arctanh\ }} \def \arcsinh {{\rm arcsinh\ }} 
\begin{document}
%%%%%%%%%%%%%%%%%%%%%%%%%%%%%%%%%%%%%%
\overfullrule=0pt
\parskip=2pt
\parindent=12pt
\headheight=0in \headsep=0in \topmargin=0in \oddsidemargin=0in
\vspace{ -3cm}
\thispagestyle{empty}
\vspace{-1cm}
\rightline{Imperial-TP-AT-2011-3}
\rightline{NSF-KITP-11-071}
\begin{center}
\vspace{1cm}
{\Large\bf  %Three point 
On strong-coupling  correlation functions  \\
\vspace{.2cm}
of circular Wilson loops  and local operators }
\vspace{1.2cm}

{Luis F. Alday$^{a,}$\footnote{alday@maths.ox.ac.uk } and Arkady A. Tseytlin$^{b,c,}$\footnote{Also at Lebedev  Institute, Moscow. tseytlin@imperial.ac.uk }}\\
\vskip 0.6cm
{\em 
$^{a}$ Mathematical Institute, University of Oxford, 
   Oxford OX1 3LB, U.K.  \\
\vskip 0.08cm
\vskip 0.08cm $^{b}$ Blackett Laboratory, Imperial College,
London SW7 2AZ, U.K. \\
\vskip 0.08cm
\vskip 0.08cm
$^{c}$ Kavli Institute for Theoretical Physics, \\
University of California,  Santa Barbara, CA 93106, USA}
\vspace{.2cm}
\end{center}

\vspace{.4cm}
\begin{abstract} 
Motivated by the  problem  of understanding 3-point correlation functions of  gauge-invariant operators in 
$\N =4$   super Yang-Mills theory  we consider  correlators  involving 
  Wilson loops  and a  ``light''   operator  with fixed quantum numbers.  At leading order in the strong coupling 
expansion   such correlators are  
given by  the ``light''  vertex operator  evaluated on a   semiclassical string world  surface  ending on the 
corresponding loops  at the  boundary of $AdS_5 \times S^5$. 
We study in detail the example of  a   correlator  of two concentric  circular Wilson loops 
  and a  dilaton vertex operator. 
 The resulting  expression  is given by  an integral of combinations of elliptic functions 
and  can be computed analytically in some special  limits. 
We also consider a generalization of the 
 minimal surface ending on two circles  to the case of non-zero angular momentum $J$  in $S^5$ and discuss
 a special  limit when one of the  Wilson loops is effectively  replaced by a 
``heavy'' operator with   charge $J$. 
\end{abstract}

\newpage
\setcounter{equation}{0} 
\setcounter{footnote}{0}
\setcounter{section}{0}

\setcounter{equation}{0} \setcounter{footnote}{0}
\setcounter{section}{0}
\vfuzz2pt % Don't report over-full v-boxes if over-edge is small
\hfuzz2pt % Don't report over-full h-boxes if over-edge is small

\setcounter{footnote}{0}

%%%%%%%%%%%%%%%%%%%%%%%%%%%%SEC1%%%%%%%%%%%%%%%%%%%%%%%%%%%%%%%%%%%%%%%%%%%%%%%%

\renewcommand{\theequation}{1.\arabic{equation}}
\setcounter{equation}{0}

\section{Introduction}

The strong-coupling 
limit of  a 3-point correlation function  of ``long'' primary operators 
 in dual  planar  $\N=4$ super YM  theory  should 
correspond, in the ``semiclassical'' approximation, to 
a  correlator of three   closed string vertex operators in \adss  string theory
 each carrying large quantum numbers 
of order of string tension $T= {\sql \ov 2\pi}$.
 This  brings  to light a challenging   problem
 of 
  how 
to construct minimal surfaces in \adss  that  effectively 
   ``end'' 
 on  three  distinct points 
at the boundary of the Poincare patch of $AdS_5$.
While  this  general problem  \ci{jan}  still awaits its solution,
progress in understanding   3-point  correlators at strong
 coupling 
was achieved \ci{za,cos,rt,bt2,gv}  in a special limit when  only two  of  the  three 
operators carry large charges
 (i.e. are ``heavy'',  $V_H$)  while the 
third one has a fixed charge (i.e. is ``light'',  $V_L$), i.e.  for the correlators of the type 
$\lan V_H V_H V_L \ran$.
  In this case the corresponding semiclassical 
minimal
 surface has cylindrical topology (i.e. is a sphere with 2 punctures) 
and is the same one that 
saturates the 2-point correlator  $ \lan V_H V_H  \ran$  of the two ``heavy'' operators \ci{t03,bu1,jan,bt2,tsu}. 
A similar idea can be applied also to the case of higher-point functions with only two ``heavy'' operators
$\lan V_H V_H V_L...V_L \ran$ \ci{rt,bt2}.

Before trying to generalize to the interesting 
case of $\lan V_H V_H V_H \ran$ 
one may step back and ask  if a  correlator with  just  {\it one} ``heavy'' operator, e.g., 
$\lan V_H V_L V_L \ran$,     may be also  computed by using a similar 
 semiclassical  approximation  at  large $\sql$, at least for some  types of ``heavy'' operators.
 At first sight,  the answer to  this question is not obvious, 
as ``heavy'' string vertex operators are naturally associated with classical string solutions which in turn 
describe a  semiclassical approximation to a  2-point function  $\lan V_H V_H \ran $. 
However, instead of  considering a cylindrical world surface one  may  start with  a  disc-like 
(euclidean) surface   and  think of  specific  
 boundary conditions that it  should end on a 
contour $C$  at the boundary of AdS  as   representing a particular  closed string state. 
Extracting an on-shell (primary-operator) 
state  is,  of course,    non-trivial:
specifying  a contour $C$  at the boundary 
%(or a Wilson loop in the gauge-theory language) 
will, in general, correspond to an off-shell string state  describing 
an  infinite superposition of local operators. 

To begin  with, one may  just   consider a simpler problem  by 
 replacing    a ``heavy''  operator by a boundary state represented by  some 
 Wilson loop $W[C]$,  i.e.  by  replacing    the question about $\lan V_H V_L V_L \ran$ by 
the one about $\lan W[C]  V_L V_L \ran$.  The semiclassical surface  for the latter  correlator 
will then be the same as the one   that determines the strong-coupling   expectation 
value  of  $ W[C]  $  (as, e.g.,  in  some  basic examples 
 discussed in \ci{malda,corr}). 

More  generally, one  may be interested in  minimal surfaces 
that  end on  several  different closed contours  at the boundary, representing
  correlators of  Wilson loops  (WL's)  and local operators   in the dual gauge theory.  
Expanding in size of   a contour,   a   WL    may be viewed, in the OPE sense \ci{corr,zar99},  as 
an infinite  sum of  local operators (and vice versa). 
 Thus this   problem is, at least  in principle,  
 related to that of  correlators involving several ``heavy''     vertex operators.

With the eventual  goal    to  understand how to  compute generic 3-point correlators  or to 
construct   minimal 
surfaces ending  on  3 separate  contours, here we shall 
consider (as in the  vertex operator case   \ci{za,rt})   an   
``intermediate''  case  of a correlator of one or  two  ``large'' 
Wilson loops  and one ``light'' vertex operator.
%\foot{Since $W[C]$ 
 %does not, in general represent a primary vertex operator, 
%the ``2-point'' correlator $\lan W[C]  V_L  \ran$  will be, in general, non-zero.} 
In this case  the semiclassical    surface  will be  determined just by the  WL's.
Our  aim    will be to study such   correlators with  the WL's being the simplest circular ones.
%\foot{Examples
% when $C$ is built out of  null lines will be considered in \ci{bt3}.} 

%%%%%%%%%%%%%%%%%%%%%%%%%%%%%%%%%%%%

Let us first discuss   the  case of  a  correlator  $ \lan W[C] V(x) \ran   $  of one 
 WL  and  a local operator. It    can  be 
 represented  by a  string path integral 
over  disc-like surfaces  (with boundary conditions specified by the 
contour $C$) of  
a  string vertex operator  $V(x)$ labelled by a point $x^\m$ of  the  
boundary of the Poincare patch of 
$AdS_5$.\foot{In our notation  the  $AdS_5$  metric  is 
$ds^2 = { 1 \ov z^2} ( dz^2 + dx^\m dx_\m )$.
Dependence of the correlator on $C$ and $x^\m$   may  be 
(partially) determined  by the conformal 
$SO(2,4)$ symmetry considerations (see below); for example, in the case when $C$ is a circle 
we may set $x_\m$ to be at 0 or at infinity.} 
If the operator is ``light'', $V=V_L$,  i.e.  it does not carry a large charge
of order of string tension 
(so that its  dimension  does not depend on $\sql$ 
   in the BPS case or grows at most as  $(\sql)^{1/2}$ in the non-BPS case) 
 it  can be ignored in finding the stationary surface in 
the path integral. Then  
the leading contribution to the normalized correlator 
$ \lan  W[C] V_L(x) \ran  \ov \lan W[C] \ran  $   will be  given just   by  the value of $V_L(x)$ on 
 the same minimal surface that determines   $\lan W[C] \ran $.
An example was considered in \ci{corr,z02}   where the WL was  a circular one \ci{go,corr,dgo} 
and the vertex operator was a BPS one with an 
angular momentum $J$  in $S^5$.\foot{The resulting surface  is  given by a  ``semisphere''  $z^2 + x^2=R^2$ 
   ending  at $z=0$ on a circle  
with the massless  bulk-to-boundary 
propagator connecting its point to a point $x$ at the boundary.
The  leading contribution at large $x$ 
is then  given   by  the corresponding OPE coefficient $c$, i.e.
%(we  assume that $C$ is a circle located at  $x=0$): 
 $ ({ \lan  W[C] V_L(x) \ran   \ov \lan W[C] \ran  })_{|x| \to \infty}
 = { 1 \ov N} {c \ov |x|^{2 J}} $.
$c$ is   proportional to the string tension  
that enters the normalization \ci{corr} of the vertex operator, i.e.  
$c= \ha \sql \sqrt{J}$.
In \ci{z02} the first  subleading contribution of 
string fluctuations near that  surface  was  shown to  give 
an extra factor  $\sim$ exp$(- { J^2 \ov 2 \sql})$. 
The derivative terms in the  chiral primary vertex operator  and the  fermionic terms in the 
string action were ignore but their  contribution should  
conspire to cancel  due to the  marginality of the vertex operator 
 and  supersymmetry.}

%AAT
In the case when  $J$ is  of order $\sql$, i.e.  the corresponding vertex operator is ``heavy'', 
  its contribution (proportional to $z^J \rX^J$ where 
$\rX=X_1 + iX_2= \cos \theta \ e^{i \vp} $  represents  an  $S^5$ rotation 
 plane) 
is to be taken into account in determining  the stationary surface in the string path integral. 
The resulting  euclidean  surface found in \ci{z02}  (see also \ci{myy}) 
is a  simple  modification of the    ``semisphere''   to include  a 
non-trivial background for the $S^5$ angle  $\vp= i \J \tau$, \ $ \J = { J \ov \sql}$, 
   which  describes the back reaction  to a ``source'' provided 
by the ``heavy'' vertex operator insertion.\foot{Here $\tau$ is the  euclidean  world sheet time coordinate,  with the boundary circle ($x^2 =R^2$) 
parametrized by  the angular  coordinate  $\phi =\sigma \in (0, 2\pi)$. 
In this paper we use (somewhat unconventionally) 
 $\s$ to parameterize the boundary contours and $\tau$ to describe  the evolution of the surface 
into the bulk. This is natural  if one thinks of a loop $C$    as representing a closed-string boundary  state rather 
than  a  trajectory   followed by the ends of an open string attached to the boundary. We shall mostly consider euclidean  surfaces in euclidean $AdS_5 \times S^5$.} 
One is also to modify the equation for $z$ to take into account 
the  contribution from the vertex operator (reflecting its scaling dimension).\foot{The  semiclassical contribution
to the correlator  is  
$  c \sim e^{ - S_{cl.}}, \ \ S_{cl.}= - \sql [  \sqrt{ \J^2 + 1} -1 +
 \J \ln (\sqrt{ \J^2 + 1} -\J) ]$.
  Here 
   $ S_{cl.}( \J \to 0) =
\ha  \sql \J^2 + ...$    reproduces  the ``subleading'' contribution in the 
 case of small $J$,  i.e. when $V$ is a  ``light'' operator \ci{z02}.}  

One may also  consider  other ways  of adding  angular momentum $J$ to a WL surface.
%an alternative (and inequivalent) way  of ``adding   angular momentum'' 
%to   a   WL world surface \ci{tz}. 
In the case  when $J$ is carried by  the vertex operator the corresponding 
R-charge  is  localized at the single  point $x$, while   one may also 
spread it over the WL contour.
One suggestion  \ci{tz}   to include rotation in $S^5$ is by choosing the unit 6-vector 
as $\theta^m=(\cos \J \s, \sin \J\s,0,0,0,0)$  (corresponding to $\vp=  \J \s$)  
 thus  getting complex scalar  coupling 
terms  $ e^{-i\J \s} Z + e^{i\J \s} \bar Z$ in the exponent of the supersymmetric  WL.  In this case
 there will be a local  R-charge density, but the total R-charge will be zero.  Another possibility  is to 
  smear a  non-zero R-charge  along $C$  by introducing $Z^J$ factors  along the contour, as, 
e.g.,  in $W[J, C] = \Tr  \big[ {\cal P}
 \int d \s  \ Z^J(x(\s))\ \exp  {\cal A}\big]$ where ${\cal A}=\int d \sigma 
  \big[ i A_\m (x(\s)) x'^\m(\s) 
  + | x'(\s)| \Phi_m  (x(\s))  \theta^m (\s) \big]$.
%N
Yet another (presumably equivalent to leading order)  option is to introduce  a linear combination of scalars 
 including a $Z$-term directly into the exponent $\cal A$.\foot{We are grateful to N. Drukker  for an important 
    clarifying discussion of    the issue of adding R-charge  to a Wilson loop.} 
In this case the $J$-dependence of  corresponding minimal surface will 
still be described by $\vp= i \J \tau$ 
but  there will be  no  modification of the dilatation charge;  
  the surface will  still be  $z^2 + x^2 = R^2$ 
 but the  expression  for $z(\tau)$  and  the induced metric  
determining $ \lan W[C, J]\ran $  will be different than in the $ \lan W[C]V_J(0)\ran $ case. 
One may then compute  $\lan  W[C,J]\  V_L(x)\ran$  by  evaluating $V_L$ on this  surface
and,  again, the result will be different from that  for  $\lan  W[C] V_J(0) V_L(x)\ran$.

%%%%%%%%%%%%%%%%%%%%%%%%%%%%%%%%%
Similar considerations apply    (as in   \ci{rt,bt2})  to the  correlators  involving\ 
(i) one WL (with or without $J$)  and several ``light'' vertex operators 
and\   (ii)  one WL, {\it one}  ``heavy'' operator and several ``light'' vertex operators.
Their leading-order semiclassical  expression will  be  given by the 
product of the ``light'' vertex operators  evaluated on the minimal surface determined by (i) 
the  corresponding WL or (ii)  by the WL and  the ``heavy'' operator.
% (or  by 
%the   generalized  Wilson loop  $W[C,J]$  with an extra semiclassical   angular momentum   $J$).
% where  $J$ stands for the  corresponding semiclassical charge. 

\

%%%%%%%%%%%%%%%%%%%%%
A correlator  of  {\it two}  WL's   and a number of ``light'' operators, 
$\lan  W[C_1] \ W[C_2]\ V_L(x_1) ... V_L(x_k) \ran$,   may   be evaluated  in a similar way, provided
 the stationary surface ending on the  two contours is known. 
The latter is  known in the case of $C_1$ and $C_2$   being   concentric  circles 
in $AdS_5$ \ci{zar99,oz}  and  this is the case which  we will consider in detail below. 
As  was  pointed out  in \ci{df}, the two circles  can be put,  by a conformal transformation,  into one plane. Furthermore, such minimal surface can be generalized to the presence  of 
an angular momentum  $J$  in $S^5$.
The latter case  may be interpreted  as corresponding to the  correlator 
$\lan  W[C_1,J]\  W[C_2,J]\ran$. 
One may also  consider   the case  when  $J$ is introduced 
instead by an insertion of a  local ``heavy'' operator,  $\lan  W[C_1] \ W[C_2] V_J(x) \ran$; 
the corresponding semiclassical  surface should be easy to construct (following the same logic as in \ci{z02})
in the special angular symmetric  case  of $V_J$ inserted 
 at the center  $x=0$ of the circles  (or at $x=\infty$, which is  related by inversion).

%The resulting  surface may also  be interpreted   as 
%providing the leading semiclassical approximation to  the correlator 
%$\lan W[C_1] W[C_2] V_H^{(J)} (0) \ran $    of two concentric 
%circular WL's and  one ``heavy'' operator $V_H^{(J)}$  inserted at the center of the circles. 
%$x=0$. % (or at $x=\infty$). 

A feature of   3-point correlation functions $\lan V(x_1) V(x_2) V(x_3)  \ran $ 
 of  primary  local operators in conformal field theories  is that their  dependence on the 
positions of the operators is fixed by conformal symmetry. When considering correlation functions
 involving Wilson loops, such as  $\lan W[C_1] W[C_2] V(x)  \ran$, the presence of the loops
 makes conformal invariance much less restrictive, so the dependence on the position
$x^\mu$  of the vertex operator  will be much richer. It is one of the aims of this paper 
to compute explicitly this dependence in several  special cases.

%%%%%%%%%%%%%%%%%%%%%%%%%%%%

This paper is organized as follows. In section 2  we  consider the simplest case of a
correlator  of a  single circular Wilson loop with a ``light'' dilaton 
 operator. In section 3  we 
study  the correlation function of a dilaton with two concentric circular Wilson loops.

In section  4  we    generalize  the discussion of section 3 
 %the classical solution corresponding to the  two concentric Wilson loops
 to the 
 case  of  the  WL's   carrying  non-zero $S^5$ angular momentum $J$.
This configuration admits two interesting limits.  In the first  limit  (section 4.2) 
 one of the two  Wilson loops shrinks 
to zero size,  and   the corresponding  surface then  corresponds to  a correlator 
of a Wilson loop with a ``heavy'' operator  of  charge $J$ inserted at the position of the 
shrunk loop. 
In the second  limit  the two circles coincide, i.e. 
 we obtain a single circular Wilson loop  carrying  angular momentum $J$.
 
Section  5  contains  some concluding remarks.
 Appendix A is devoted to  the study of the action of conformal symmetries on the  correlators 
 considered in this paper.
In Appendix B we compute the  correlator of section 4 in the  second special limit mentioned above.

%%%%%%%%%%%%%%%%%%%%%%%%%%%%%%%%
\renewcommand{\theequation}{2.\arabic{equation}}
\setcounter{equation}{0}

\section{Semiclassical correlation function  of  circular Wilson loop\\  and dilaton operator 
%mputation of 3pt functions
}

As discussed  in the introduction,   Wilson loops play   a  similar   role  to that  of  ``heavy'' operators  
in a  correlator   with ``light'' operators -- at large $\sql $  they determine the semiclassical trajectory
on which  the   ``light'' operators   should be evaluated to get the leading strong-coupling contribution. 
% in order to computed the three-point function of a light operator with two heavy operators (or in our case, two Wilson %loops), it suffices to evaluate the light vertex operator on the classical solution of the two heavy operators. 

In this paper we will  consider   configurations 
 embedded into the  $AdS_3 \times S^1$ subspace of  (euclidean)  \adss  with metric
\begin{equation}
ds^2=z^{-2}(dz^2+dx_\mu dx^\mu  )+   d S_5 = z^{-2}(dz^2+dr^2+r^2 d\phi^2	+ ...)+d\varphi^2 + ... \ ,  \la{as}
\end{equation}
where $x_1=r \cos \phi,~\  x_2=r \sin \phi$.
% span the plane in which the solutions are embedded. 
As ``light'' operator we will consider  the dilaton vertex operator 
\bea
&&   V_L (x' ) = \int d^2 \zeta \  U\big(x(\zeta) - x', z(\zeta), \vp(\zeta) \big)  \ , \la{opa}  \\
&&U(x,z,\vp) =k_{\Delta}\  \Big( \frac{z}{z^2+x^2} \Big)^\Delta\ e^{ i  j \vp} 
\   {\cal L}  \ , \ \ \ \ \
{\cal L} =   z^{-2}(\partial z \bar \partial z+\partial x_{\mu} \bar \partial x^{\mu})
+\partial \varphi \bar \partial \varphi  + ...  \no
\eea%\la{opa} 
Here $j$ is an $S^5$ momentum  and $\Delta = 4 + j$ is the marginality condition. 
  For  simplicity 
we shall suppress the 
 normalization factor  \ci{corr,za,rt}\ \   $k_{\Delta}= { \sql \ov 8 \pi N } \sqrt{ (j+1)(j+2) (j+3) } $ 
 in  most of  the expressions for the correlators 
  in this paper.
Dots  stand for  terms with   other $S^5$  coordinates and   fermions 
that  will not be relevant at the leading semiclassical approximation considered below. 
The operator is labelled by a  point $x'$  at the boundary ($z=0$)  of the Poincare patch.

In what follows we shall 
 parametrize the location of the vertex operator by $(x'_1,x'_2)=\rho(\cos \theta,\sin \theta)$ 
and by  $h$ --  the transverse distance in the $(x'_3,x'_4)$ plane, so that 
\be
z^2 + (x-x')^2 =  z^2  + r^2   -2 r \rho\ \cos(\phi-\theta)+ \rho^2 +h^2     \ . \la{jop}
\ee
Note that in the case of the    dilaton operator integrated over the position of the insertion point $x'$, i.e. 
$\int d^4 x' \ V(x')$, one  is  to do the replacement  
\be
\Big( \frac{z}{z^2+(x-x')^2} \Big)^\Delta \rightarrow
 \int d^4x'\  \Big( \frac{z}{z^2+(x-x')^2} \Big)^\Delta=\frac{z^{4-\Delta}}{2(\Delta-1)(\Delta-2)} \ . \la{nnl}
\ee
This factor  is  trivial only  if $j=0$, i.e. when  $\Delta=4$. 

In the  case of a correlator  involving one circular WL  the corresponding  classical solution can be represented  in the 
form ($w,\psi$ are  the radial and angular world-sheet  disc coordinates in conformal gauge) 
 \ci{corr,dgo}
 %(
%The simplest solution, of the type considered in this paper, corresponds to a single Wilson loop. 
%Denoting the world-sheet coordinates by $\zeta= w e^{i \psi}, \bar \zeta= w e^{-i \psi}$, 
%the solution is simply
\bea
&&z^2 + r^2 = R^2  \ , \ \ \ \ \ \    \vp=0 \ , \la{ci}\\
&&z=R \frac{1-w^2}{1+w^2},~~~r=   \frac{2 Rw}{1+w^2},~~~~~~~~\phi=\psi\ , \ \ \ \ \  \ 
\zeta= w e^{i \psi} \ , \ \ \    0 \leq w \leq 1  \ . \la{cii}
\eea
%\la{cii}
This surface ends on a circle of radius $R$ at the boundary $z=0$. 
%as we approach the boundary, we approach a circular Wilson loop of radius $R$.
Let us  note that under the inversion symmetry of the AdS metric \rf{as} 
(which becomes the standard  inversion at the boundary  $z=0$) 
\be \la{inv}
 z\to { z \ov z^2 + x^2} \ , \ \ \ \ \ \ \  \ \  x_\mu \to { x_\mu \ov  z^2 + x^2} \ , \  \ \ \ \ \ \ 
z^2 + x^2 \to { 1 \ov z^2 + x^2} \ , \ee
this  surface  goes into a similar one with $R \to 1/R$. 

The leading semiclassical expression for the  correlator of one circular WL  with one ``light'' 
dilaton operator  is  then be given by 
\bea 
&&\C
%_{(W[C],V_L)}
\equiv  { \lan  W[C]\ V_L (x') \ran  \ov \lan  W[C]  \ran } =    \big[V_L(x') \big]_{semicl.} \no \\
&&=
\frac{8\pi k_{\Delta}}{\Delta-1} \ \frac{R^{\Delta}}{\left[(h^2+\rho^2-R^2)^2+4 R^2 h^2\right]^{\Delta/2}}
\ , \la{sop}
\eea
%\la{sop}
where we plugged in the  solution \rf{cii} into \rf{opa},\rf{jop}  and performed the 2d integral
(the dependence on $\theta$  dropped out due to  angular  symmetry of the solution). 
%adially symmetric, hence the dependence on $\theta$ will drop out. 
This agrees 
with the  expression  found in \cite{corr}. 
A salient feature of this expression  is that it diverges as $d^{-\Delta}$  when  the distance $d=\sqrt{ (\rho-R)^2+h^2}$ between 
the vertex operator  insertion point $x'$  and the
Wilson loop becomes small. In  Appendix A we show how one could 
find  this result by using the conformal group  symmetry considerations.
% of a single circular  Wilson loop.

In the following sections we will consider several other  examples of  solutions 
involving circular Wilson loops. The  resulting expression for  the correlators $\C$ 
will be much more  involved  being   given in terms of an integral of elliptic functions, but
% which cannot be solved in general.
we will be able to analyze several limits of interest.

%%%%%%%%%%%%%%%%%%%%%%%%%%%%%%%%%%%%%%%%%%%%%%%%%
\renewcommand{\theequation}{3.\arabic{equation}}
\setcounter{equation}{0}

\section{Correlation function of  two concentric  circular Wilson loops\\ and a dilaton operator }

In this section we will consider the
 3-point correlation  function of two concentric  circular Wilson loops 
and  a  light dilaton vertex operator,  i.e.  $\langle W[C_i] W[C_f] V_L (x)  \rangle$, 
where $C_i$ and $C_f$ stand for an ``initial'' and  ``final''  circles 
on which the semiclassical surface will be ending. 
 
\subsection{Classical solution for surface  ending on  two concentric circles}

Let us start  with the   semiclassical solution 
corresponding 
to two concentric circular Wilson loops \ci{zar99,oz}.  Using conformal transformations
one may  arrange  the 
 two concentric circles  to   lie  in the same plane
at the boundary  \cite{df} (see also appendix A).\foot{The solution we discuss below 
is  equivalent to the one in  \ci{oz,df}; note that  our notation  are   different from those in \ci{df}.} 
In this section we restrict ourselves to the  case without rotation  in $S^5$ ($\varphi=0$), i.e. 
the  surface will lie  only     inside  of $AdS_3$ 
% The form of the solution is
\begin{equation}
z=z(\tau)\ ,~\ \ \ \ ~~r=r(\tau)\ ,~~\ \ \ \ \ \ ~\phi=\sigma\ , \la{kl} 
\end{equation}
where  $\tau$ (ranging from $\tau_i=0$ to $\tau_f$) 
 and $\sigma$ (having   period $2\pi$)  are   coordinates of a 2-d cylinder.
% (we use conformal  gauge).
 The boundary conditions are 
\begin{equation}
z(0)=z(\tau_f)=0 \ ,~~~\ \ \ r(0)=R_i \  ,~~~\ \ \ r(\tau_f)=R_f\ , \la{jjj} 
\end{equation}
where $R_{i},R_f$ are the radii of the  two circular Wilson loops. 
Using the  conformal gauge we  have the usual constraint  or vanishing of the 
 ``2d energy'' (dot will denote derivatives with respect to $\tau$)
\be 
  z^{-2} (\dot z^2 + \dot r^2  - r^2 )= 0  \ , \la{hoh}
\ee
 and also the  following  integral of motion
(dilatation charge)
\begin{equation}
z^{-2}(z \dot z +r \dot r )=p \ , \la{pap}
\end{equation}
where $p$ is a constant  that  will   parametrize the solution. 
From the space-time point of view, the solution is parametrized by the ratio $R_i/R_f$, so there 
is a relation (given below)  between this ratio and $p$. 
Note also that under the inversion symmetry \rf{inv} 
the equation \rf{pap}  goes into itself with $p \to - p$, 
i.e. the  solutions with $p >0$ and $p < 0$ are related  by the inversion. 
%AAT
The special case of $p=0$  corresponds to the case when the two  radii are the same, $R_i=R_f$, 
i.e. the two circles coincide and we get back to the ``semisphere'' surface. 

It is convenient to perform the following change of variables:
\begin{equation}\la{pai}
z=\frac{u e^v}{\sqrt{1+u^2}},~~~\ \ \ \   ~r=\frac{e^v}{\sqrt{1+u^2}}\ . 
\end{equation}
In terms of $u(\tau), v(\tau)$ the  integrals of motion  \rf{hoh}
 and    \rf{pap} become 
\begin{equation}
\dot u ^2=1+ u^2 -p^2 u^4 \ ,~\ \ \ \  ~~~ \dot v = { p u^2 \ov u^2 +1} \ . \la{uu}
\end{equation}
The boundary conditions imply that $u \rightarrow 0$ and $e^v \rightarrow R_{i,f}$ at the two 
 boundary circles.  
It is useful to introduce  the following combinations (the roots of 
the polynomial $1+ u^2 -p^2 u^4$)
\begin{equation}
u_{\pm}^2=\frac{1 \pm \sqrt{1+4 p^2}}{2p^2} \ . \la{ko}
\end{equation}
For  real solutions 
(having $p^2\geq 0$)  only $u^2_+$ is positive. 
The first equation in \rf{uu} 
 can be solved \ci{df}  in term of  the  elliptic integral  $F(y|m)$  of the first kind
 expressing  $\tau$ as a function of $u$
\begin{equation}
\tau = u_+ F \Big(\arcsin \frac{u}{u_+} | \frac{u_+^2}{u_-^2} \Big) \ . \la{tyy}
\end{equation}
 $u$ has a maximum (a turning point) at $u_+$.
 Beyond  it, we have to continue on the other branch of the  arcsine function, until reaching the boundary again. 
The full range of $\tau$ (or $\tau_f$, since we have set $\tau_i=0$) is given by 
\begin{equation}
\tau_f= 2 u_+ K \Big(  \frac{u_+^2}{u_-^2}\Big) \ , \la{yyy}
\end{equation}
where $K(x)$ is the complete elliptic integral of the first kind. 

One can then    find  $v(\tau)$ in terms of elliptic integrals. 
Its expression, as that of $\tau(u)$, will depend on whether we are
 on the first or the second brach. For the first branch ($u$ from $0$ to $u_+$) 
\begin{equation}\la{rrr}
\tau = u_+ F \Big(\arcsin \frac{u}{u_+} | \frac{u_+^2}{u_-^2} \Big),~~~\ \ \ \ \ \ \ \
v=v_i+\hat v(u)
\end{equation}
while for  the second branch ($u$ from $u_+$ to $0$) we obtain
\begin{equation}\la{ttt}
\tau =  \tau_f -u_+ F \Big(\arcsin \frac{u}{u_+} | \frac{u_+^2}{u_-^2} \Big) \ , 
~~~ \ \ \ \ \   v=v_f-\hat v(u) \ , 
\end{equation}
where
\begin{eqnarray}
&&\hat v(u)=p u_+ \Big[ F \Big(\arcsin \frac{u}{u_+} | \frac{u_+^2}{u_-^2} \Big)- \Pi \Big(-u_+^2,\arcsin \frac{u}{u_+} | \frac{u_+^2}{u_-^2} \Big)\Big] \ , \la{li} \\
&&v_f-v_i  = \log \frac{R_f}{R_i} = 2 p u_+ \Big[K \Big(\frac{u_+^2}{u_-^2} \Big)-\Pi \Big(\-u_+^2,\frac{u_+^2}{u_-^2} \Big) \Big]\ . \la{lii}
\end{eqnarray}%\la{lii}
The latter  equation  thus gives the  expression for the modulus of the solution $R_f/R_i$ in terms of  the integral of motion 
$p$.

Let us now analyze the space-time structure of this solution. For definiteness, we can take $v_i=0$,  i.e. $R_i = e^{v_i} =1$.
 We can then plot $v_f$  in  \rf{lii}  as a function of $p$, see fig. 1.
%\iffalse
%%%%%%%%%%%%%%%%%%%%%%%%%%%%%%%%%%%%%%%%%%%%%%%%%%%%%%%%%
\begin{figure}[h]
\centering
\includegraphics[scale=0.5]{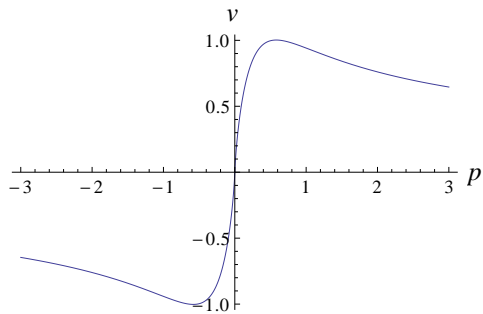}
\caption{The  relation between $p$ and  the ratio of radii 
$R_f/R_i=e^v$. Note that this ratio  is bigger  than  1  
 for $p > 0$  and smaller than 1  for $p < 0$. Furthermore, in general,  there are 
 two different values of $p$ corresponding  to the same   value  of $v$ or    $R_f/R_i$.}
\end{figure}
%%%%%%%%%%%%%%%%%%%%%%%%%%%%%%%%%%%%%%%%%%%%%%%
%\fi 
From this plot we can infer  that $p>0$ corresponds to $R_f=e^{v_f}>1$, while $p<0$ corresponds to $R_f<1$. 
As was mentioned below \rf{pap}, 
the two  configurations are related by an inversion. 
We may focus, e.g.,   on  the case of $p>0$. 
The radius $R_f  > 1 $ cannot be  chosen  arbitrarily large 
since there is a value $p =p_0 \approx 0.5811$  for which $R_f=e^{v_f}$ reaches a maximum, $R_{0}$.
This indicates 
 a phase transition  \ci{go,dgo,zar99,oz}, in which  two separate 
 minimal surfaces, one per each Wilson loop, dominate over this solution. 
Note also    that  given  some  value of  $R_f$, \  $1 < R_f < R_{0}$
 %($R$ is greater than one, since $p$ is positive), 
there are two values of $p$ corresponding to it   (see fig.1 ). 

In order to understand how these solutions look like in space-time
 we can go back to the original coordinates $(z,r)$   in  \rf{pai} and plot $z(r)$  parametrically,
 varying $u$  (see fig. 2). 
%using $(r,z)=\frac{e^{v(u)}}{\sqrt{1+u^2}}(1,u)$.
 For instance, both $p=0.0244$ and $p=35.52$ correspond to $v_f=0.2$. 
%\iffalse
%%%%%%%%%%%%%%%%%%%%%
\begin{figure}[h]
\centering
\includegraphics[scale=0.4]{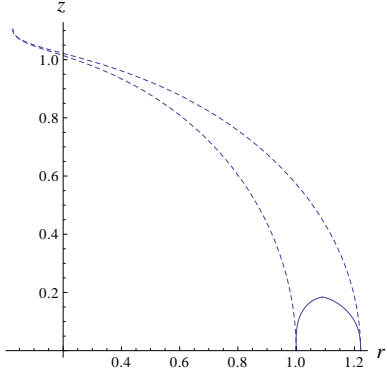}
\caption{Two different solutions with the same value of $R_f/R_i$. 
The solution corresponding to $p<p_0$ represented  by the dashed line  
 tends to the single Wilson loop solution as $R_f/R_i \rightarrow 1$.}
\end{figure}
%%%%%%%%%%%%%%%%%%%
%\fi
The dashed  curve 
 corresponds to $p=0.0244<p_0$. It starts at $z=0, r=R_i=1$, then 
 goes up to some $z=z_{max}$  almost reaching  $r=0$ and then back to the outer Wilson loop, 
at $z=0, r=R_f \approx 1.2$.
 Note that in the limit when 
 the inner Wilson loop approaches the outer (which corresponds to very small $p$), this world-sheet surface  approaches 
 the surface  for a single Wilson loop   \rf{cii}.
 The solid curve corresponds to $p=35.52>p_0$. Note that
it disappears in the limit $R_f \to 1$  %which is  consistent with the range  of $z$ 
  %$0 < z < z_+$, as $z_+$  tends to zero in this limit.
as then the  range of variation of  $z$   shrinks  to 0. 

We can also study  what happens when $p$ is closer to  the special value 
 $p_0$, see figure 3. 
%\iffalse
%%%%%%%%%%%%%%%%%%%%%%%%%%%%%%%%
\begin{figure}[h]
\centering
\includegraphics[scale=1]{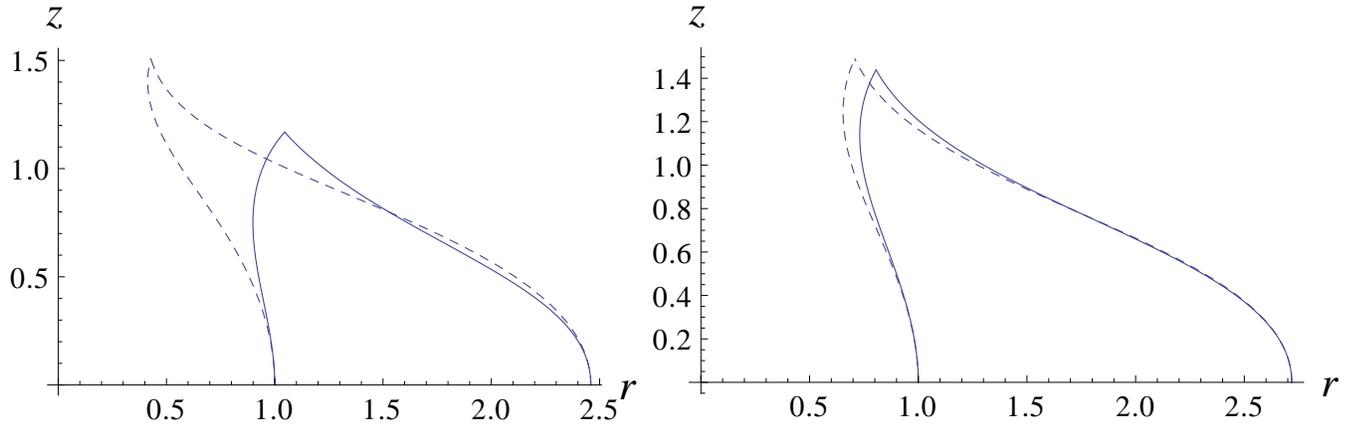}
\caption{The two different solutions approach each other as the corresponding  two values 
of $p$  get closer to $p_0$.}
\end{figure}
%%%%%%%%%%%%%%%%%%%%%%%%%
%\fi
Here again  the dashed line corresponds to $p<p_0$ 
while the solid line corresponds to $p>p_0$. We see that both  curves 
%the blue and red line 
reach a minimal  value of $r$, where $\dot r(u)=0$. This value happens at $u=1/p$ 
and is given in terms of elliptic functions. 
An important feature is that this minimal  value increases monotonically with $p$.

To conclude,   in general,  we get two distinct solutions for a given value  of    $R_f/R_i$. 
 One can evaluate the euclidean string action on both of them; the action  turns out to be 
negative after the subtraction of a divergence \ci{oz}.  The solution with 
$p > p_0$ representing non-intersecting surface (solid curve)  has bigger action  and thus  should   dominate
over  the solution  with $p  < p_0$  (dashed curve). A related remark  appeared in \ci{oz}. 

In  doing the semiclassical computation for a correlator   with a ``light''  vertex operator  
we are, in general,   supposed to sum over all   competing stationary points, i.e.
\bea 
&& \C=  { \lan  W[C_i ] \  W[C_f]\  V_L(x)  \ran  \ov \lan  W[C_i] \  W[C_f]  \ran } =   
{   b_1 e^{ \sql a_1}   + b_2 e^{ \sql a_2} + ...   \ov  c_1 e^{ \sql a_1}   + c_2 e^{ \sql a_2} + ...  } \no \\
&& = {b_1 \ov c_1}   +  \Big(  {b_2 \ov c_1}  - {b_1  c_2\ov c^2_1}  \Big) \   e^{ - \sql (a_1-a_2) } + ... \ . 
\la{tak}
\eea%\la{tak}
 Here we assumed that $a_1 > a_2$  and $b_k \sim \sql$ (taking into account the normalization of the vertex operator
in \rf{opa}). 
While only one (dominant) solution  will contribute to the leading  term in \rf{tak}, 
below    we shall formally   consider the case of a general $p> 0$, i.e. 
we will  treat  the cases of the two solutions on an equal footing.

%%%%%%%%%%%%%%%%%%%%%%%%%%%%%%%%%%%%%%%%%%%%%%%
\subsection{Semiclassical correlation  function }

To compute the leading  semiclassical   correlation  function \rf{tak} 
we need to evaluate the  dilaton vertex operator \rf{opa} on the  classical solution described 
in the previous subsection. 
 It is convenient to express the integral over the cylindrical 
world-sheet coordinates  $(\tau,\sigma)$  using $u(\tau)$ instead of $\tau$. 
%in terms of the functions $u(\tau),v(\tau)$. Furthermore, since $u(\tau)$ is given as an inverse function, it is convenient to %express the integral as an integral over $du$ rather than over $d\tau$. 
This will introduce the following factor  in the measure (see \rf{uu})
\begin{equation}
\dot \tau(u)= { d \tau \ov d u} = \big[ (1-\frac{u^2}{u_-^2})( 1-\frac{u^2}{u_+^2})\big]^{- 1/2} \ . \la{ks} 
\end{equation}
Here we have chosen the sign plus  for the first branch; 
the world-sheet integral is an integral over the two branches but the different sign in the measure 
will be  compensated by the interchange of  the limits of integration. 
We end up with the following expression for the correlator in \rf{tak}\foot{Here and below
 we suppress the normalization factor $k_{\Delta}$ of the dilaton operator 
in the expressions for the correlators.} 
\begin{eqnarray}
& &\C =   \big[V(x) \big]_{semicl.} =
I(R_i)+I(R_f) \ , \la{juu} \\
I(R_{i,f})&=&\int^{2 \pi}_0 d\sigma\int^{u_+}_0 
 du\  \frac{2 \dot \tau(u)}{u^2} \Big( \frac{u\ e^{v_{i,f} \pm \hat v(u)}}{\sqrt{1+u^2}\big[e^{2(v_{i,f}\pm \hat v(u))}+h^2+\rho^2\big] -2e^{v_{i,f}\pm \hat v(u)}\rho \cos \sigma } \Big)^{\Delta} \nonumber
\end{eqnarray}
where we used \rf{jop}  and   eliminated the dependence on $\theta$ 
by a shift of $\sigma$  (the dependence on $\theta$ drops out due to angular symmetry).  
%where the $u$ integral runs from zero to $u_+$.

The integrand depends on four external parameters: $v_{i,f}= \log R_{i,f}$ (i.e. 
the radii of the Wilson loops)  and $\rho,h$  (the location of the vertex operator).
It is useful to study  how $AdS_5$ 
symmetries act on this expression. For instance,  if one   acts with dilatations,
\begin{equation}
h \rightarrow   \ell  h,~\ \ \ ~~\rho \rightarrow   \ell \rho,~\ \ \ 
~~v_{i,f} \rightarrow \log \ell  + v_{i,f} \ , \la{oouu}
\end{equation}
then the integrand   picks up  a factor  $ \ell^{-\Delta} $.
%$\C \rightarrow   a^{-\Delta} \C$. 
This allows  us to set,  e.g.,  $v_i=0$ or $R_i=1$. 
Then $\C$ will depend on three dimensionless parameters, i.e. on 
 $p,\rho,h$  (and,  of course, on $\Delta$). 
If one  acts  with the  inversion (cf. \rf{inv}) 
\begin{equation}
h \rightarrow \frac{h}{h^2+\rho^2}\ ,~~\ \ \   ~\rho \rightarrow \frac{\rho}{h^2+\rho^2}\ 
,~~\ \ \ \ ~v_{i,f} \rightarrow - v_{i,f},~\ \ \  ~~p \rightarrow -p\ , \la{tre}
\end{equation}
then  one can explicitly check  (using that $\hat v(u)$ changes sign when $p$ changes sign)   that  the integrand picks up  a 
factor $(\rho^2+h^2)^{-\Delta}$.

%We would like to compute this integral. Unfortunately,
The integral over $\s$ can be done using that 
\bea \la{inn}
 \frac{1}{\pi}\int_0^{2\pi} d\sigma \  \big(\cos \sigma+s  \big)^{-\Delta}
=(s-1)^{-\Delta}~_2F_1\big(\ha,\Delta,1,{\te \frac{2}{1-s}}\big)+(1+s)^{-\Delta} ~_2F_1\big(\ha,\Delta,1, {\te \frac{2}{1+s}}\big) \eea 
where we have assumed $s>1$. The integral diverges for $s \rightarrow 1$, which will happen when the insertion point of the operator approaches one of the Wilson loops.

The remaining  integral over $u$ is  hard to compute  explicitly
 due  to the  dependence of the integrand  on $\hat v(u)$ given   in \rf{li}. 
 Still, it can be  easily evaluated numerically  for any 
choice of the external parameters. 
Furthermore,  it can be, in principle,   computed analytically  in  special  limits
that  we discuss below. 

\subsubsection{Operator   close to  a Wilson loop}

An interesting limit corresponds to the case   when  position of the vertex operator 
is close to one of the boundary circles. 
In this case 
% take the insertion point very close to one of the Wilson loops. In such a limit,
 we  should expect the  correlator  to look like the one  for the 
 single Wilson loop   discussed in section 2. 
Since  the distance to one of the concentric  loops (say $C_f$) is 
$d^2=(\rho-R_{f})^2+h^2$, we expect  to find that  for $d \to 0$  the correlator  diverges   as 
% of the form
\begin{equation} 
\C \approx \frac{2^{2-\Delta}}{\Delta-1} d^{-\Delta}  \ .\la{ki}
\end{equation}
A numerical evaluation of \rf{juu}  confirms this 
expectation. This result can also be understood directly from \rf{juu} by noticing that the divergence comes from the region close to the boundary $u \approx 0$. There, and for $d \approx 0$, $s$ in \rf{inn} approaches one.

%%%%%%%%%%%%%%%%%%%%%%%%%%%%
\subsubsection{Small $p$ limit}

 Small $p$  limit is possible only  for the first ``self-intersecting'' 
solution  (which, as was mentioned above,  is, in general,  subdominant). 
 In this limit   the classical solution approaches the one for a single circular Wilson loop, 
i.e. $R_f \to R_i$. 
For  small  $p$ we have
\begin{equation}
\hat v(u) = v^{(1)}(u) \ p\  +{\cal O}(p^3),~~\ \ \ \ \ \ \ 
~v^{(1)}(u)= {\rm arcsinh} \ u -\frac{u}{\sqrt{1+u^2}} \ . 
\end{equation}
Let us adopt  here  the symmetric choice $v_f=-v_i=\ha \delta v $  (i.e. $R_f = 1/R_i $).
 For small $p$    (i.e. $R_f = 1/R_i \to 1$)
\begin{equation}
\delta v = 2 \big[ \log (p/8)   +1 \big] \ p\ +{\cal O}(p^3) \ .
\end{equation}
The integral over $u$ is from 0 to $u_+$, with $u_+ \to \infty$ for $p \to 0$.
Performing  the integrals in \rf{juu}   up to $u_+$ and then expanding  the result in powers of $p$  we obtain
for the correlator 
\be
\C=\C^{(0)}+p^2 \C^{(2)}+{\cal O}(p^4),~~\ \ \ \ \ \ \ \
~ \ \C^{(0)}=\frac{8\pi}{\Delta-1} \frac{1}{\big[(h^2+\rho^2-1)^2+4 h^2\big]^{\Delta/2}} \la{klp} \ . 
\ee
As expected, $\C^{(0)}$  matches   the expression \rf{sop}  
for the correlator of a  dilaton  operator and a single  circular 
  Wilson loop of unit radius.
For 
 $\C^{(2)}$  we find 
% is harder to compute, since the information about $\hat v(u)$ enters at this order. We obtain
\begin{eqnarray}  &&
\C^{(2)}=-4\pi b^{-2\Delta}+2b^{-2\Delta}f_1(y)+ 2b^{-2\Delta-4}\big[
4(1-b^2)+\Delta(b^2-2)^2\big] f_2(y)+4 b^{-2\Delta} f_3(y) \ ,\no  \\
&& \ \ \ \ \ \ \ \ \ \   b^2\equiv h^2+\rho^2 +1   \ , \ \ \ \ \ \  y\equiv { \rho \ov  h^2+\rho^2+ 1 }  \ ,  \la{kpp}\\
&& f_1(y)= \int du d\sigma \frac{u^{2+\Delta}}{(1+u^2)^{3/2}(\sqrt{1+u^2}-2 y \cos \sigma)^\Delta} \ , \la{ooo}\\
&&f_2(y)= \int du d\sigma \frac{\Delta \ u^{\Delta-2}\sqrt{1+u^2}}{\big(
\sqrt{1+u^2}-2 y \cos \sigma\big)^{\Delta+2}}\ \big[v^{(1)}(u)+1+\log(p/4)\big]^2 \ , \\
&& f_3(y)=\int du d\sigma\  y \cos \sigma \frac{\Delta \ u^{\Delta-2}}{(\sqrt{1+u^2}-2 y \cos \sigma)^{\Delta+2}}\ \big[v^{(1)}(u)+1+\log(p/4)\big]^2 \ . 
\end{eqnarray}%\la{kpp}
%where $b^2=1+h^2+\rho^2$ and $x=\rho / b^2$. 
For  $\Delta=4$ the functions $f_k(y)$ entering $\C^{(2)}$ simplify %slightly 
 and  can be computed, for instance,   as an  expansion in powers of $y$
(i.e. in the limit of small $\rho$ or large $h$). 
%Their expressions, however,  are  quite lengthy and not very iluminating.

%%%%%%%%%%%%%%
\subsubsection{Large $p$ limit}

In a similar way, one
 may also  compute $\C$ for  large values of $p$.
This  limit is possible only for the  dominant non-self-intersecting solution.
It corresponds to  a 
``small" world-sheet surface  between two  close circles with radii approaching 1. 
%Wilson loops which are very close. 
%The easiest way to compute the expansion is to
It is useful to  change the variable $u \to \tilde u = { u \ov u_+}$
%, \ \   $ to $u \rightarrow \tilde u u_+$, 
so that  $\tilde u$ runs from zero to one. 
As in the previous case, we shall 
set $v_f=-v_i=\ha \delta v$. 
%We can expand the fuctions entering the integrand for large values of $p$. We obtain
For large $p$  we   find  (here $E$ and $F$ are the elliptic integrals) 
\bea\la{hhp}
&&\delta v=2\pi^{1/2}\frac{\Gamma[\fot]}{\Gamma[\fo]}\ p^{-1/2}+...\ , \\
&&\hat v(u)=\Big[E(\arcsin\tilde u,-1)-F(\arcsin \tilde u,-1)\Big] \ p^{-1/2}\ +... \ , 
\eea%\la{hhp}
and at the  leading order  we get 
\begin{equation}
{\cal C}^{(0)} =p^{1/2-\Delta/2}\frac{\sqrt{\pi}\Gamma[\fo 
(\Delta-1)]}{\Gamma[\fo (\Delta+1)]} \int_0^{2\pi} d\sigma 
\frac{1}{\big(1+h^2+\rho^2-2\rho \cos \sigma\big)^\Delta}\ , \la{uyu}
\end{equation}
which can easily be solved in terms of hypergeometric functions using (\ref{inn}).  
This result is in  good agreement with the numerics, see figure 4.

\begin{figure}[h]
\centering
\includegraphics[scale=0.5]{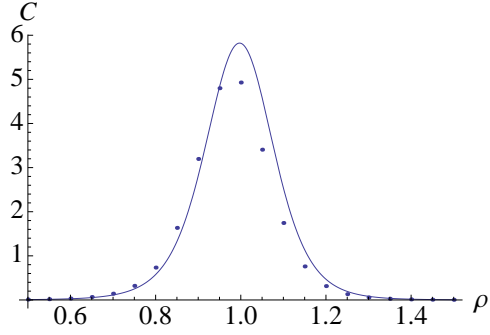}
\caption{Analytic expression for the correlator
$\C(\rho)$  in the large $p$ limit (solid line) versus numerical result (dots). We have chosen a large value of $p$ and $h=0.2$ in order to smooth out the singularity at $\rho=1$.}
\end{figure}

We see that the result is smooth except for a singularity at $h^2+(\rho-1)^2=0$. This singularity is,  of course, 
 expected, since there we approach the location of the Wilson loops.
 By looking at the singularity of (\ref{inn}) as $s \rightarrow 1$ we obtain that in this limit
\begin{equation}
{\cal C} \approx \big[ h^2+(\rho-1)^2\big]^{-\Delta + \ha }\ . 
\end{equation}
This is a milder singularity  than  in the case of a single Wilson loop (\ref{ki}). 
This is to be expected, since in this limit the world-sheet also disappears.

Finally, let us mention that, with some effort, the subleading terms can also be computed.
Explicitly,   for  $\Delta=4$ we get
for the first correction to \rf{uyu}  ($\C= \C^{(0)}  + \C^{(1)}  + ...$) 
\begin{eqnarray}
 \C^{(1)}=p^{-5/2}\frac{128 \sqrt{2\pi}b^2}{3(b^2-4\rho^2)^{11/2}(\Gamma[\fo])^6}  \Big(
\pi^2 \big[ b^8(5-5b^2+b^4)+b^4(100-100 b^2+23 b^4)\rho^2 \no 
\\ +6(25-25 b^2+8 b^4)\rho^4+6 \rho^6\big]
%+ \nonumber \\
+ 640 (\Gamma[{\textstyle{5 \ov 4}}]) ^8 (b^{12}+14 b^8 \rho^2-25 b^4 \rho^4-96 \rho^6) \Big) \ , \la{gjj}
%\nonumber
\end{eqnarray}%\la{gjj}
where $b^2=h^2+\rho^2+1$   as in \rf{kpp}. 

%%%%%%%%%%%%%%%%%%%%%%%

%%%%%%%%%%%%%%%%%%%%%%%%%%%%%%%%%%%%%%%%%%%%%%%%%%%%%%
\renewcommand{\theequation}{4.\arabic{equation}}
\setcounter{equation}{0}

\section{Case of circular Wilson loops with   angular momentum in $S^5$}

For the classical solution  ending on two concentric circles  discussed in the previous section 
 the ratio of radii $ { R_i \ov R_f} \equiv  { R_{inner}\ov R_{outer}} < 1 $  cannot be made  arbitrarily small --
there is a  critical value after  which   the  dominant surface is  a  combination of disjoint   
``cups''   ending   separately on each of the two circles.
Here we shall  consider, following \ci{df}, a generalization  of the above solution to the 
case of non-constant angle of $S^1$ in $S^5$, 
which, after continuation to Minkowski signature, may be interpreted as an $S^5$ 
angular  momentum  carried by the Wilson loop. 
 In this  case  there will be a nontrivial $ { R_i \ov R_f}  \to 0$  limit  of 
the corresponding surface  which will be equivalent to  the surface   \ci{z02} 
which  ends  on  the ``outer'' circle and is also   ``sourced''  by a  ``heavy''
vertex operator with a semiclassical angular momentum   $J$.

%AAT
%%%%%%%%%%%%%%%%%%%%%%%%%%%%
\subsection{Classical solution  and general form of the correlator   with   dilaton operator}

We start with the following  generalization \ci{df}   of the ansatz  in \rf{kl} 
\begin{equation}\la{ann}
z=z(\tau),~~~~~r=r(\tau),~~~\phi=\sigma,~~~~~~\varphi= i \J\tau  \ . 
% \ , \ \ \ \  \ \ \    a=i \J \ .
\end{equation}
\iffalse
An equivalent euclidean surface  will be found if we  choose instead 
\be \varphi=  \J\s  \ , \la{qoj} \ee
corresponding to the WL carrying angular momentum  $J= {\sql}\J $  in the sense of \ci{tz}. 
The choice  of  $\varphi= i \J\tau$ is  more natural in order to relate to the surface 
for a circular WL and a vertex operator with $J$ in a special limit when one of two circles shrinks to zero 
that we  discuss below.
\fi
%We  assume that $a$ can be  imaginary, with $J= {\sql}\J $ interpreted as an 
%angular momentum   after  the continuation  $\tau \to  i \tau$. 
The conformal constraint \rf{hoh} is then  replaced by 
\be 
z^{-2} ( \dot z^2 + \dot r^2 - r^2) = \J^2  \ ,\la{kop}
\ee
while the second integral of motion  is the  same as in \rf{pap}.  Then  the first equation in \rf{uu} 
also keeps a similar   form:
\be \dot u^2=1+(1+\J^2) u^2 -(p^2-\J^2)  u^4  \ .\la{kf} 
  \ee
In this section  we will consider the case in which $p^2-\J^2>0.$\footnote{A special case where 
$p^2-\J^2<0$  will be  considered in Appendix B.}
 In this case the resulting solution  is analogous  to the one with  $\J=0$, e.g.,
 the structure of the world-sheet surface, with   two branches, is exactly the same.
 The only difference is that now $u_{\pm}$ in \rf{ko} should be replaced by
\begin{equation}
u_{\pm}=\frac{1+ \J^2 \pm \sqrt{(\J^2-1)^2+4 p^2}}{2(p^2-\J^2)} \ . 
\end{equation}
The resulting expression for $R_f/R_i$ in terms of $p$ and $\J$ is then 
\begin{equation}\la{rr}
\log \frac{R_i}{R_f} =- 2 p u_+ \Big[K \Big(\frac{u_+^2}{u_-^2} \Big)-\Pi \Big(\-u_+^2,\frac{u_+^2}{u_-^2} \Big) \Big]\ . 
\end{equation}
We   now have an  interesting possibility of taking
 $ p^2 - \J^2  \rightarrow 0$  (so that $u_{\pm}$ become very large), while keeping $p$ finite
which corresponds to ${R_i \ov R_f }\to0$. 
In this   limit $p=\pm  \J $  the solution  becomes  equivalent to the one found in \ci{z02};
written in terms of $(u,v)$ in \rf{pai} it  takes  the following  simple  form 
\be
\label{limsol}
u=\frac{1}{\sqrt{1+p^2}}\sinh(\sqrt{1+p^2}\ \tau)\ , \ \ \ \ \ \ 
v=p \tau-\arctanh \Big[ \frac{p}{\sqrt{1+p^2}}\tanh(\sqrt{1+p^2}\ \tau) \Big] \ .
\ee
 In this limit the configuration of  two circular 
 Wilson loops degenerates to that of a single Wilson loop plus an effective local  operator, which is located at the 
position of the ``shrunk''  loop, i.e. at 0 
 for $p<0$ or at infinity for $p>0$  (the two cases are, of course, related by  an  inversion). 
 This  is a ``heavy'' operator  $V_H^{(J)} $  parametrized by $p=\J$  
 which    has an  interpretation of a  semiclassical  angular momentum
$J = \sql \J \gg 1  $.
 The space-time geometry can be visualized  by plotting $(z,r) $ for different values of $p$,
for $\tau \in  (0,\infty)$. If $p$ is positive, then the solution starts at $(r,z)=(1,0)$,
where the Wilson loop is located, and ends at $(0,\infty)$, where the operator is located.
 If $p$ is negative, the solution ends at $(0,0)$ instead, i.e.  the operator is located at the origin. 
%We can also see that  for $p$ negative, $z$ vanishes as $ e^{p \tau}$ as $\tau$
 %becomes large, so $p$ plays the role of the $R-$charge of the operator.

Another special case corresponds to $p=0$ for $\J\not=0$. 
In this case  \rf{pap}  implies that  $z^2 + r^2 = R^2 =\const$ or $v$ in \rf{uu} is  constant. That means 
$R_i=R_f$, i.e. the two circles coincide  and the form of the surface is again the ``semisphere''.
At the same time,  $z(\tau)$ or $u(\tau)$ in \rf{pai}    is non-trivial as       \rf{kf} 
becomes $\dot u^2=1+(1+\J^2) u^2  + \J^2   u^4$. This case is equivalent to that of a single circular WL with an addition of  
angular momentum $J$ and will be considered in detail in Appendix B. \foot{
Note that 
while in the first limiting case the angular momentum  density is singular as $1\ov r$, i.e. $J$ is concentrated 
near $r=0$ where the operator is located, in the second limit  the density is $1\ov \sqrt{1-r^2}$
which is integrable, i.e. $J$ is spread along the surface (cf. discussion  in the introduction).} 

Let  us now consider the  correlation   function  of two  generalized ($\J\not=0$) 
   circular Wilson loops and a 
dilaton vertex operator \rf{opa}  (located at $(\r,h)$) 
with  fixed $S^5$ angular momentum $j$, i.e. with  $\Delta=4+j$. 
The integrand of the correlator simplifies if we use the first-order constraints \rf{pap},\rf{kop} satisfied by $z(\tau)$ and $r(\tau)$. 
After introducing the variables $u(\tau),\ v(\tau)$ as in \rf{pai}  we end up with the following expression
for the corresponding analog of  \rf{juu}\foot{
%N
In \rf{ann}  we have  chosen the  initial condition for $\varphi$ so that it vanishes  at $\tau=0$. 
Keeping $\vp_0=\vp(0)$ non-zero will introduce an extra phase factor $e^{i \vp_0 j}$ in 
$\C$.}
\begin{equation}
\label{cint}
\C=  \int^{2 \pi}_0 d\sigma \int^{\tau_f}_0 
d\tau\ \frac{2}{[u(\tau)]^2} e^{i a j \tau} \Big[\frac{e^{v(\tau)} u(\tau)}{
(e^{2v(\tau)}+h^2+\rho^2)\sqrt{1+u^2(\tau)}  -2 e^{v(\tau)} \rho \cos \sigma} \Big]^\Delta \ .
\end{equation}
In the following section we shall study this correlator in the special case (\ref{limsol}).

%%%%%%%%%%%%%%%%%%%%%%%%%%%%%%%%%%%%%%%%%%%%%%%
\subsection{Special case of one degenerate  circle ($p=\pm \J$)   }

Let us  consider  the  special  case   of $ p=  \pm  \J $  (we shall assume that $\J > 0$).
 For $p >0$ the radius of the  ``outer'' circle goes  to infinity. 
For  $p<0$
 the ``inner''  circle   shrinks to zero, i.e.  is effectively replaced by a
local operator $V_H^{(J)}$.
 After some simplifications, we get the following 
expression for the corresponding correlators  (below we set $R_f=1$)
\begin{eqnarray}
&&\C^{(\pm)} = { \lan   W[C_f]\  V_H^{(J)}\  V_L^{(j)} \ran  \ov \lan   W[C_f]\  V_J \ran  }
%_{(WV_HV_L)}
=2 \la{kou} 
\int^{2\pi}_0  d\sigma \int^\infty_0 
d\tau\  e^{(\Delta\pm j) p \tau}\ \frac{1+p^2}{\sinh^2(\sqrt{1+p^2}\ \tau)}  \no  \\
&&
\times \Big[
\frac{\sqrt{1+p^2}\tanh g(\tau) - p }{h^2+\rho^2+e^{2p \tau}(1+2p^2)-
2e^{p \tau}  \sqrt{1+p^2} \  
\big[\cosh g (\tau)\big]^{-1}   \big[\rho \cos \sigma+p e^{p \tau}   \sinh g(\tau)\big]      } \Big]^{\Delta}\ ,   \no \\
&&g(\tau)\equiv \sqrt{1+p^2}\tau+\arcsinh p \ . 
\end{eqnarray}
The   $\pm$    choice  comes from the  sign in  $p=\pm \J$.
% (we shall assume that $\J > 0$). 
% We could have also written the answer as an integral over the variables $u$ and $\sigma$. This will introduce a factor %$\tau'(u)=\frac{1}{\sqrt{1+(1+p^2)u}}$ in the measure.
Since  under the inversion $p \to - p$,
$\C^{(+)}$ and $\C^{(-)}$  should  also  be  related by an inversion 
transformation.  Indeed, as one  can show  explicitly, 
\begin{equation}
\C^{(\pm) }\big(p,\Delta,\rho,h\big)   = 
(h^2+\rho^2)^{-\Delta}  \ \C^{(\mp)}\big(-p,\Delta,\frac{\rho}{\rho^2+h^2},\frac{h}{\rho^2+h^2}\big) \ . 
\la{kyy}
\end{equation}
The integrand  in  \rf{kou} depends on the  momentum $j$  of  the  ``light''  dilaton  operator  ($\Delta=4+j$), 
%, which are related if we satisfy the marginality condition $\Delta=4+j$,
 on $p$ or  the momentum  $J$  of 
the effective  ``heavy''  vertex  operator,
 and on  the location of the dilaton 
 operator, parametrized by $\rho$ and $h$.
%  (the  dependence on the angle $\theta$ in \rf{jop} again dropped out due to 
%rotational symmetry of the surface). 

%As the radius of the Wilson loop is set to one, $C$ does  not transform anymore covariantly under dilatation. On
% the other hand, 
Below we will compute the correlators \rf{kou} in several special limits.

%%%%%%%%%%%%%%%%%%%%%%%%%%%%%%%%%%%%%%%%%%%%%%%%%%%%%%%

%%%%%%%%%%%%%%%%%%%%%
\subsubsection{Small $p$ limit}

The small $p$ limit is very simple but interesting nonetheless.  In this limit the ``heavy''
 operator parametrized by $p$ becomes   effectively ``light''
 and the solution (\ref{limsol}) tends to the circular Wilson loop solution plus a wiggle going from the center of the world-sheet to the origin, see figure 5.

\begin{figure}[h]
\centering
\includegraphics[scale=0.4]{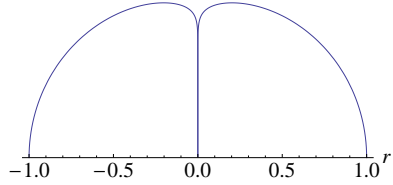}
\caption{$z(r)$ plot  of surface in  the small $p$ limit:  we obtain a single Wilson loop with an  insertion of a  ``light'' operator.}
\end{figure}
In this limit the dependence on  $j$ is not seen, 
and we simply obtain, as expected,  the same result as in the  single circular Wilson loop case.

\subsubsection{Large $p$ limit}

%$C$ can also be computed for large values of $p$. 
Taking $|p|$ large at  leading order we get  
%(here $\k=\pm 1$)
\begin{equation}
\C^{(\pm)}
= |p|^{\pm j-1}2^{\pm j+1} \int^\infty_0  du\int^{2\pi}_0 d\sigma \ \frac{u^{\Delta
\pm  j-3}}{\left(1+h^2+\rho^2+u^2-2\rho \cos \sigma \right)^{\Delta}}\ . \la{kppp}
\end{equation}
Performing the integrals  gives 
\begin{eqnarray}
\C^{(\pm)} &=& \pi |p|^{ \pm j-1} 2^{\pm  j+1} \ (1+h^2+\rho^2)^{\ha ( \pm j-2-\Delta)}\ 
\frac{\Gamma[\ha (2+\Delta\mp  j)]\ \Gamma[\ha (-2+\Delta\pm  j)]}{\Gamma[\Delta]} \nonumber \\
& &\ \  \times ~_2F_1\Big(\fo (2+\Delta\mp  j),\fo (4+\Delta\mp  j),1,\frac{4\rho^2}{
(1+h^2+\rho^2)^2}\Big)\la{kqq} \ . \la{mm} 
\end{eqnarray}
These results are in perfect agreement with the numerics. Using the marginality condition   $\Delta=4+j$,  we obtain
\begin{eqnarray}
&&\C^{(+)}= 2^{2+j}\pi |p|^{j-1} \frac{\Gamma[j+1]}{\Gamma[j+4]} \ \frac{(1+h^2)^2+2(h^2+2)\rho^2+\rho^4}{\big[ (1+h^2)^2+2(h^2-1)\rho^2+\rho^4\big]^{5/2}}\ , \la{lk}\\
&&\C^{(-)}= 2^{1-j} |p|^{-j-1}\pi \frac{(1+h^2+\rho^2)^{-j-3}}{3+j}~_2F_1\big(\frac{3+j}{2},\frac{4+j}{2},1,\frac{4\rho^2}{(1+h^2+\rho^2)^2}\big) \ . \la{fhg}
\end{eqnarray}
% \la{fhg}
 %finding perfect agreement.
Note that $\C^{(\pm)}$ diverges in the limit  $h^2+(\rho-1)^2\to 0$ 
since the argument of the hypergeometric function becomes one. This is, 
 of course,
 the expected divergence as  then the location of the vertex operator approaches  the Wilson loop
 (of radius $R=1$).

%%%%%%%%%%%%%%%%%%
\subsubsection{OPE limits}

%A17
When the location of the dilaton  operator approaches the remaining finite   Wilson loop, 
we expect a  singularity  that is the same as in the case of 
 a single Wilson loop \rf{ki}, i.e.
 %After all, the details of the solution everywhere else should not matter. This divergence goes as
 $d^{-\Delta}$, where $d$ is the distance to the Wilson loop.
This is readily confirmed  by a numerical evaluation of the integral \rf{kou}.
% Numerically we find perfect agreement with this expectation.

Another  interesting
 limit is when the location of the dilaton operator becomes very close to
  the location of the degenerate    circle. Namely, close to the 
location of the effective  ``heavy''
 operator, which is at the origin for $p<0$ (the case we consider below). 
%(the case when the ``heavy'' operator  is at infinity is related by an inversion).
This happens in the limit $d =\sqrt{h^2+\rho^2} \ll 1$.\foot{The correlator $C^{(+)}$ 
is regular in this
limit  (it is singular for $d \gg 1$).}
In this case we find the leading singularity to be 

\be
 \C^{(-)} \ \sim \ \pm\  \  {f(p,j) \over  d^{\g }}  \  ,\ \ \ \ \ \ \ \ \
\g = 
4+ 2j + 2  \sqrt{1+p^{-2} }    \ ,  \la{yyo}\ee 
\be f(p,j) =-4\big[ 1+2(p^2-|p|\sqrt{1+p^2})\big]^{\ha (j+\sqrt{1+p^2})}\ {(1+p^{-2})} \ 
\frac{\Gamma[2-{\sqrt{1+p^{-2}}} ]\ \Gamma[2+j+{\sqrt{1+p^{-2}  } }  ]}
{\Gamma[4+j]} \la{saa} \no
\ee
%This  expression is  singular at  $d\to 0$ for a range of values of $p$ and $j$ 
%on the sign of $\g_\pm$ which in turn depends on 
%the value of $p$.
\iffalse
\bea
&& \C^{(\pm)} \ \sim \ \pm\  \ f(\pm p,j) \   d^{-\g_\pm }  \  ,\ \ \ \ \ \ \ \ \
\g_\pm \equiv  
4+ j(1 \mp 1 )+ 2 p^{-1} \sqrt{1+p^{2} }    \ ,  \la{yyo}\\
&& 
f(p,j) =-4\big[ 1+2p(p-\sqrt{1+p^2})\big]^{\ha (j+\sqrt{1+p^2})}\ {1+p^{2} \ov p} \ 
\frac{\Gamma[2-{\sqrt{1+p^{2}}\ov p} ]\ \Gamma[2+j+{\sqrt{1+p^{2}  } \ov p}  ]}
{\Gamma[4+j]} \la{saa} \no
\eea
This  expression is  singular or regular for $d\to 0$ depending on the sign of $\g_\pm$ which in turn depends on 
the value of $p$.
\fi
It would be interesting  to understand the interpretation of 
the power $\g$.  On general grounds,  one 
%While naively one 
 may expect  this limit to be related to an OPE  involving the dilaton operator
$V^{(j)}_L(x)$  and   an effective ``heavy'' operator $V^{(J )}_H(0)$ and indeed, the leading contribution to (\ref{yyo}) comes from the region close to the heavy operator, or large $\tau$ in (\ref{kou} ), 
see \ci{za} for a related discussion.

%%%%%%%%%%%%%%%%%%%%%%%%%%%%%%%%%%%%
\subsubsection{Large $j$ limit}

Assuming $j$ and thus $\Delta=4+j$   are large  and considering for simplicity the case when the dilaton operator is located at the origin  ($h=\rho=0$) 
we get from \rf{kou}
\bea
&&\C^{(\pm)}=\int^\infty_0  d \tau\  e^{\pm j p \tau} \frac{2(1+p^2)}{\sinh^2(\sqrt{1+p^2}\ \tau)}
\ \Big[\frac{e^{ q(\tau) }  \ \sinh(\sqrt{1+p^2}\ \tau)}
{\sqrt{1+p^2+\sinh^2(\sqrt{1+p^2}\ \tau)} }
 \Big] ^{4+j}
\ , \la{gh}\\
&& q(\tau)\equiv 
-p \tau + { \rm arctanh\ } \big[{ p\ov \sqrt{1+p^2}}   \  \tanh(\sqrt{1+p^2}\ \tau) \big] \ .
\la{qw}
\eea
For  $p >0$, we find that 
%The integral hence depends on $p$ and $J$. It turns out that
 for large $j$ the integral  $\C^{(+)}$ grows exponentially
 %A17
 %, while $\C^{(-)}$ 
%decays  %, with the following behavior (remember $p>0$). 
\be
\C^{(+)} \approx    j^{-1-\frac{2p}{\sqrt{1+p^2}}} \  { \rm exp } \Big( {-\frac{1}{2}\log[
 1+2p(p-\sqrt{1+p^2})] \ j}\Big) \ .
   \la{hh} \ee
%&& \C^{(-)}  \approx  j^{-1/2} \    { \rm exp } \Big(
%-  \big[{2 p\ov \sqrt{1+p^2}} \  \arcsinh \frac{\sqrt{1+p^2}}{\sqrt{2} \ p } \big] \ j    \Big)   \ . \la{qqt}
%\end{eqnarray}
%
%Again, these results are in perfect agreement with the numerics.

%%%%%%%%%%%%%%%%%%%%%%%%%%%%%%%%%%%%%%%%%

%A17

\subsubsection{Correlator with integrated dilaton operator}

%{\bf below $p$ stands for $|p|$ ??} 

If we consider the  case of the correlator   with the dilaton operator integrated over the 
insertion point  (see \rf{nnl}) 
then it turns out that  $\C^{(-)}_{int}$  can be computed  explicitly as a 
 function of $p$ and $j$
%($\C^{(+)}_{int}$ appears to be  more complicated):
\begin{equation}
\C^{(-)}_{int}=\int_0^\infty du\  \frac{2u^{j-2}}{\sqrt{1+(1+p^2)u^2}}\left(
1+\ u\ \big[u+2p^2u+2p\sqrt{1+(1+p^2)u^2}\ \big] \right)^{-j/2} \ .\la{kpq}
\end{equation}
This integral is convergent for $j >1$. 
The  easiest way to compute it  is by  first expanding in powers 
of $p$ and  then integrating term by term. For the first few terms we get
\begin{equation}
\C^{(-)}_{int} =\frac{2}{j-1}-2p+(j-1)p^2-{1\ov 3} j(j-2)p^3+...\ . \la{iii}
\end{equation}
One can then guess the general form of these  terms to be 
$(-1)^n2^n \frac{\Gamma[{1 \ov 2} (j+n-1)]}{\Gamma[{1 \ov 2}  (j-n+1)]\Gamma[n+1]}p^n$, and 
finally  resum the  series. 
 We arrive at the following simple result  valid for all values of $p$ and $j>1 $
\begin{equation}
   \C^{(-)}_{int} =\frac{2}{j-1}\left(p+\sqrt{1+p^2} \right)^{1-j} \ ,\ \ \ \ \   j > 1  \  .\la{hq} 
\end{equation}
For   $j=0$  the  insertion into a correlator of the integrated dilaton operator \rf{opa},\rf{nnl}  
is equivalent to  an  insertion of the string action   and thus  to  a   differentiation  of the  correlator 
over the string tension.
% (which  at the leading semiclassical  order  brings  down the classical string action 
%evaluated on the solution).   For $j=0$ the regularized integral \rf{kpq}   
% written in terms of   $y= u^{-2}$ 
\bea 
\C_{int}(\eps) &=& \int  du\  \frac{2u^{-2} }{ \sqrt{1  +(1+p^2) u^2 }}  
=2 \int_{\tau_i}^{\tau_f} d\tau \frac{1+p^2}{\sinh^2(\sqrt{1+p^2}\ \tau)}\no \\
&=& -2 \sqrt{1+p^2} \ \coth(\sqrt{1+p^2}\ \tau) 
\Big|_{\tau_i}^{\infty} \ \to \   - \sqrt{ 1 + p^2}    \ , \la{kq} 
\eea
where  we first  changed back to the world-sheet coordinate $\tau$, introduced a 
cutoff $\tau_i$  (that should correspond to a cutoff 
in $z= \eps \to 0$) and finally dropped the singular term. This expression is the same 
(up to an overall normalization factor) 
as the   corresponding ``bulk'' string action  in    eq. (4.15) of \ci{z02} 
or   eq.(3.27) of \cite{df}.

\renewcommand{\theequation}{5.\arabic{equation}}
\setcounter{equation}{0}

\section{Concluding remarks }

In this paper we  studied the leading semiclassical approximation to correlators of Wilson loops and local operators. 
One  motivation is to learn more about how to represent  local operators by (limits of) particular Wilson 
loops and  also  to shed light on the structure of 3-point correlation  functions of local operators. 
We have seen that  correlators involving two Wilson loops and a local operator have rather non-trivial dependence on the quantum numbers and position of the  operator.

It would be  useful  to generalize  our discussion to  other more complicated Wilson loops 
like  in \ci{pz}  and also  study other examples of  simple Wilson loops for which minimal surfaces are known 
explicitly, as, e.g., in  \ci{df,bur,kkk}. 

Another  particularly interesting  case is that of Wilson loops built out light-like  segments  \ci{kru,am,krut}.
With a special choice of a contour (as in  weak coupling picture \ci{kor}) 
 correlators  of such loops with  local  operators   may be used to represent 
correlators  involving  large-spin  twist 2  operators   \ci{bt3}. 
Similar  correlators may be related also to the form factor problem \ci{maz}.

More generally, it would be interesting to understand whether leading   semiclassical results for such correlators 
may be found indirectly   by  some combination of integrability-based  methods as in \ci{gai,gv}. 

\iffalse
questions:    what about 2d conf symmetry/ Mobius ? 
we divide over it and that allows us to fix 1 point  --  position 
of one operator. 
Why then we are integrating over it ?  the answer is that we are 
still integrating over the  target space conformal group -- remaining generators not fixed  by $C$ and $x$. that needs  clarification. 
We may demand that  at C small $ < W[C] V > $ reduces to $ < VV> $ and that 
is defined with division over Mobius ...
\fi

%%%%%%%%%%%%%%%%%%%%%%%%%%%%%%%%%%%%%%%%%%%%%%%%%%%%%
\section*{Acknowledgements}

We would like to thank E. Buchbinder,  S. Giombi,  G. Korchemsky,  R. Roiban  and,  especially,  N. Drukker 
  for very  useful discussions. 
  %A17
  We also thank K. Zarembo for 
  a clarifying  comment on the first version of
  this paper. 
We also thank  E. Buchbinder for  a  collaboration on a related problem. 
Part of this work was completed while AAT was visiting KITP at the University of California, Santa Barbara 
where his research was supported in part by the National Science Foundation  under Grant No. NSF PHY05-51164.
AAT  also acknowledges  the support of  the  RMHE grant 2011-1.5-508-004-016.

%%%%%%%%%%%%%%%%%%%%%%%%%
\appendix
%\addcontentsline{toc}{section}{Appendices}
%\addcontentsline{toc}{section}{Appendices}
\section*{Appendix A:   Consequences of conformal symmetry
}
\refstepcounter{section}
\def\theequation{A.\arabic{equation}}
\setcounter{equation}{0}
%\section{Appendix: Symmetries}

In this appendix we shall discuss restrictions imposed  on the correlators   we considered  in this paper 
by the $AdS_5$ issommetry group  or conformal symmetry of the boundary theory.
The  symmetries of the boundary configurations  extend to symmetries of the corresponding 
minimal surface in $AdS_5$ that ends on them.\foot{Below  we do not distinguish between 
the cases of the  Euclidean and Minkowski signature of the boundary.}

%%%%%%%%%%%%%%%%%%%%%%%%%
\subsection{Single circle}

Let us start with a circular loop 
%We start by considering which subgroup of $SO(2,4)$ preserves the circular Wilson loop, following closely \cite{Bianchi:2002gz}. Let us  the circular Wilson loop 
\begin{equation}
x_1^2+x_2^2=R^2,~\ \  \ \ \  ~~x^3=x^{0}=0\,\la{a}
\end{equation}
and identify  
which symmetries leave it invariant up to reparametrizations.
 The infinitesimal generators of the conformal group  $(P_\mu,J_{\mu \nu},D,K_{\mu})$ 
 have the following action  (with parameters  $(a^\mu,\omega^{\mu \nu},\ell,b^\mu)$) 
  on the boundary  coordinates
\begin{equation}
\delta x^\mu=a^\mu+\omega^{\mu \nu} x_\nu + \ell  x^\mu+ (x^2 \eta^{\mu\nu} -2 x^\mu x^\nu) b_\nu\  . \la{aa}
\end{equation}
 Following  \cite{Bianchi:2002gz}, let us introduce the notation $x^l=(x^1,x^2)$ and $x^t=(x^3,x^0)$
and note that along the loop the transformations become
\bea
&&\delta x^l = a^l+ \omega^{l m}x_m+\ell x^l+R^2 b^l-2 b^m x_m x^l\ ,  \\
&& \delta x^t = a^t +\omega^{t l} x_l+R^2 b^t \ .\la{kot} 
\eea
The conditions $\delta x^t=0$ and $x_l \delta x^l=0$ imply
\begin{eqnarray}
\label{singleconst}
a^t=-R^2 b^t,~~~~~~\omega^{tl}=0,~~~~~~a^l=R^2 b^l,~~~~~~~\ell=0\  .\la{s}
\end{eqnarray}
That gives  a total of $2+4+2+1=9$ constraints, leaving six transformations that leave the loop invariant. These transformations are generated by rotations in the plane of the loop, $J_{12}$, rotations in the transverse plane, $J_{30}$ and four additional symmetries 
\be \la{ji}
\Pi^+_l=R P_{l}+ R^{-1}  K_{l} \ , \ \ \ \ \ \ \ \ \  \Pi^-_t=R P_t- R^{-1}K_t  \ .\ee
 These generators form  an  $so(2,2)$ subalgebra.\foot{If we were in two dimensions, then we would have three generators, giving 
 basically by the Mobius transformations that leave a unit (after a dilatation) circle invariant.} 
One can also work out the finite action of these generators on the space-time coordinates  (see  \cite{Bianchi:2002gz}).

%%%%%%%%%%%%%%%%%%%%%%%%%%%%%
\subsection{Two   concentric circles}

To leave two concentric circular  Wilson loops invariant, we need stronger constraints.
 In particular, in (\ref{singleconst}) we should have $a^t=b^t=a^l=b^l=0$. This leaves us with two generators: 
 rotations in the plane of the loop and rotations in the orthogonal plane.
That   means,  in particular,   that a correlator   of 2 concentric  circles  with a local 
operator at  an arbitrary position   will depend only on the  radial directions   $\r$ and $h$ in the two 
orthogonal planes.

In  \cite{zar99,oz}  the correlator  of  two concentric circular  Wilson loops 
lying on  different parallel planes was    considered. This  configuration    is, in fact, 
 related \ci{df} by a conformal transformation to the configuration where  the   circles lie in the same  plane. 
To see this,  let us  start with  two concentric circles (with  radii $1$ and $R_0$)
 in the same plane  and consider the transformation  $K_{l}-P_{l}$ along one of 
the transverse coordinates $x_t$, which leaves the unit circle invariant. From the above 
expressions for the infinitesimal transformations its easy to deduce  that the 
 loop of radius $R_0$ at $x_t=0$ is mapped, by a finite transformation with parameter $b$,
 to a concentric  loop of radius $R(b)$ at $x_t(b)$ where (here we assume   that, e.g., $x_t=x_3$ and 
dot stands  for $d \ov d b$) 
\begin{eqnarray}
\dot R(b) = 2 x_t(b) R(b),~~~~~~~~~~\dot x_t(b)=1+x^2_t(b)- R^2(b) \ .\la{bab}
\end{eqnarray}
%
%where $b$ is the parameter of the finite transformation. 
These equations can be easily integrated; assuming  $R(0)=R_0,\ x_t(0) =0$  we get 
%the apropriate boundary conditions we obtain 
\begin{equation}
R(b) = \frac{2 R_0}{1+R_0^2-(R_0^2-1)\cos 2b },~~~~~~~~~~~~
x_t(b)=\frac{(R_0^2-1)\sin 2b }{1+R_0^2-(R_0^2-1)\cos 2b}\ .\la{hoj}
\end{equation}
This   gives  us two concentric circles  of radii 1 and $R(b)$  separated by $x_t(b)$.
This implies,  in particular, that  the configuration  of two concentric  circles  of equal 
radius $R=1$ separated by a distance  $d$  is equivalent to two concentric circles in one plane with radii 
equal to 1 and   
\be R_0^{\pm} = 1 + \ha d ( d \pm  \sqrt{  d^2 +4 } ) \ ,\ \ \ \ \ \ \ \ \ \
R_0^{+}  R_0^{-}= 1 \ ,\la{rad} \ee
where the two sign options are related by an inversion.

%%%%%%%%%%%%%
\subsection{Correlator of a circular   Wilson loop  and a local   operator}

Another configuration we considered  in this paper is that of a circular 
 Wilson loop  with its center at the origin 
and 
a local operator inserted  at some arbitrary point 
%A17
(see also \ci{dk}).
 It is clear that by rotations we can always bring the operator to the location $(\rho,0,h,0)$. It turns out that one can act with the generator $\Pi^+_1$ in \rf{ji}
 so that after the transformation $\rho \rightarrow 0$, i.e.
 the operator is at a new location $(0,0,h',0)$. Then, we can act with the generator $\Pi^+_3$
to achieve  that $h' \rightarrow 0$,  i.e. that 
 the operator is located at the origin. 
This shows that we can use four of the symmetries of the circular Wilson loop in order to 
bring an operator from an arbitrary   location to the origin.

Once the operator is at the origin, we can ask which generators  leave  the resulting  
configuration invariant. An infinitesimal translation of the origin simply acts as
%\begin{equation}
$\delta x^\mu  = a^\mu$
%\end{equation}
%
so that   we need to require $a^t=a^l=0$. Hence, 
again, this configuration will be left invariant by 
two transformations:  rotation in the plane of the loop and rotation in the orthogonal plane.

The fact that a local operator can always be brought to the origin by conformal transformations that leave a given circular Wilson loop invariant    suggests  that we should be able to fix the  form of their correlator 
 (\ref{sop}) using symmetry considerations.  This is indeed the case.
Let $(\rho,0,h,0)$ be the location of the operator
in the original correlator.
  According to the above discussion, a  2-parameter family of transformations $\Pi_{a,b}$
generated by  $\Pi^+_1$ with parameter $a$ and  by 
$\Pi^-_3$ with parameter $b$  that leaves the Wilson loop invariant 
 acts  on $(\rho,h)$ as
\be
\label{pitransf}
\delta \rho=R^2 a+(\rho^2+h^2)a-2 a \rho^2+2 b h \rho\   , \ \ \ \ \   \ \ \ \ \
\delta h= R^2 b-(\rho^2+h^2)b+2b h^2-2 a \rho h \ .
\ee
% 
%where $a,b$ are infinitesimal parameters. 
Naively, we would expect $\Pi_{a,b}$ to leave the  correlator  
\be \C ={ \lan W[C]\  V_L(\r,h) \ran\ov 
 \lan W[C] \ran} \ee
 invariant. This is almost the case. Translations are, of course, 
 a  manifest  symmetry of the problem,  but  conformal transformations are not:  one finds that  $\C$
transforms as 
%Actually, we find an anomaly
\begin{equation}
\label{an1}
\Pi_{a,b}:\ \ \ \ \    \log \C \rightarrow \log \C' =\log \C + 2\Delta(a \rho-b h) \ . 
\end{equation}
The   origin of this transformation 
  is due to an anomaly  under inversions, since special conformal transformations 
are obtained  by a combination of  translations $P$ and  two inversions $I$, i.e.
 $K= I P I$. Indeed, one can  check that under the inversions 
\begin{equation}
\label{an2}
I:\\ \ \ \  \log \C \rightarrow \log \tilde \C = \Delta \log \frac{h^2+\rho^2}{R^2}+\log \C \ . 
\end{equation}
It should be possible to derive these anomaly equations along the lines of
the argument of  \cite{Drukker:2000rr}  about the  anomaly in the transformation from a 
 line to a circle. 

Equations (\ref{an1}),(\ref{an2}) 
can be checked explicitly from our result for the correlator \rf{sop}. 
We can now reverse the logic and say that if  $\C$  is 
 some  function $f(\frac{\rho}{R},\frac{h}{R})$, then the transformation 
 (\ref{an1}) implies 
two first-order differential equations, that indeed  fix  the form of this  function.\footnote{See \cite{Gomis:2008qa} for a very neat symmetry argument.}

\appendix
%\addcontentsline{toc}{section}{Appendices}
%\addcontentsline{toc}{section}{Appendices}

\refstepcounter{section}
\def\theequation{B.\arabic{equation}}
\setcounter{equation}{0}
%\section{Appendix: Symmetries}

%AAT
%%%%%%%%%%%%%%%%%%%%%%%%%%%%%%%%%
\section*{Appendix B:   Case of coinciding circular Wilson loops with  $J\not=0$
 %correlator of   circular Wilson loop  with $\J\not=0$  and dilaton
% \\
% with  non-zero angular momentum
}
In this Appendix we consider a special limit of the solution (\ref{ann}) in which $p \rightarrow 0$, while $\J$ is kept finite. In this limit (\ref{pap}) implies that $z^2+r^2=R^2$, so that the surface corresponds to a semisphere, 
as for a single circular Wilson loop.  Here $v=0$ (i.e. $R_i=R_f=1$) 
 while  the equation for $u(\tau)$  in  (\ref{kf}) becomes
\begin{equation}
\dot u^2=1+(1+\J^2)u^2+\J^2u^4 \ . \la{yyu}
\end{equation}
 As discussed in \cite{df}, the structure of the solution  of this equation 
is different  from  the case of  $p^2-\J^2>0$ we  discussed in section 4.1. 
 Now the solution does not have turning points  but again  has two branches.
 On  the first branch the solution starts  at  the boundary $r=1$ and then 
  reaches the point  $r=0$;  after that we should continue to the second branch, 
in which the solution goes from $r=0$ back to the boundary at $r=1$ .
 Again,  $\tau$ can be  found  as a function of $u$. For the first and the second branch we obtain,  respectively, 
\bea
&& \tau = \tilde u_+ F \Big(  \arctan \frac{u}{\tilde u_+}\big| 1-\frac{\tilde u_+^2}{\tilde u_-^2} \Big),~~ \ \ \ \ \
\tau =\tau_f-\tilde u_+ F \Big(  \arctan \frac{u}{\tilde u_+}\big| 1-\frac{\tilde u_+^2}{\tilde u_-^2} \Big),\\ 
&& \tau_f= 2 \tilde u_+ K \Big(1-\frac{\tilde u_+^2}{\tilde u_-^2} \Big)\ , 
 \ \   \ \ \ \ \   \tilde u_+={1\ov \J} \ ,\ \ \ \ \ \ \ \ \ \ \tilde u_-=1  \ . \la{tauf}
\eea
%
 %$v(\tau)$ vanishes identically in this limit.  We have introduced $. Furthermore,
 $u$ runs from zero to infinity for the first branch and then back to zero for the second. 
The string  action  computed   on this solution is (after removing a linear divergence)  \cite{df}
\bea
&&S=- 2 {\sql }\ \frac{1}{\tilde u_+}\ E \Big(1-\frac{\tilde
u_+^2}{\tilde u_-^2} \Big)
= - 2 \sql\ \J\ E (1- \J^{-2}) = - 2 \sql\  E (1- \J^{2})  \ , \\
&& S_{\J\to 0} =  - 2 \sql  \Big[ 1 +  \J^2 ( - \ha \log \J  + \log 2
- \fo) + {\cal O} ( \J^3) \Big] \ ,\\
&&
 S_{\J\to \infty} =
  - 2 \sql  \Big[ \J  +  { 1 \ov \J}  (  \ha \log \J  + \log 2 - \fo)
+ {\cal O} ({1 \ov  \J^2})  \Big] \ . \
\eea

This  action has a  rather remarkable ``duality symmetry'' under $\tilde u_+ \leftrightarrow \tilde u_-$ due
to the elliptic function identity $E(1-{1\ov y})={1 \ov \sqrt y}
E(1-y)$.\foot{Note that for  $\J>1$ we have $\tilde u_+ < \tilde u_-$,
so that
$\tilde u_{+}$  and $\tilde u_{-}$  should  be interchanged  but that
leaves the form of the action  invariant. The same will happen with
all the expressions below, i.e.  by  our choice of $\tilde u_{\pm}$ we
cover all the values of $\J$.}
For small $\J$ the action  starts with twice the  value of the action
of the circular Wilson loop
(here we have two coinciding loops as a limit of self-intersecting
surface  and that doubles the semisphere surfaces and thus  the
action)
while for large $\J$  the action has familiar ``near-BPS''  behaviour.

To compute the correlator in this limit we need to evaluate 
 the dilaton vertex operator on  this classical solution.  Using $\Delta=4 + j $, we obtain
\bea
{\cal C} = e^{-{ 1 \ov 2}  \J j \tau_f}\int^{2\pi}_0 d\sigma \int_0^{\infty} &du& \frac{4u^{2+j}}{\sqrt{1+u^2}\sqrt{1+\J^2 u^2}} 
\no \\
&\times& \frac{\cosh\Big( j  \big[ F\left( \arctan \J u,1-\J^{-2}\right)-K\left(1-\J^{-2} \right)   \big]\Big)   }{\big[\sqrt{1+u^2}(1+h^2+\rho^2)-2\rho \cos \sigma \big]^{4+j}} \la{kwe}\ , 
\eea
where we assume that $j, \J \geq 0$.
% Note that the overall prefactor $e^{-{ 1 \ov 2}  \J j \tau_f}$ can be absorved in a redefinition of the origin of $\tau$, so we drop %it for the rest of the appendix. 
% $\kappa=\pm 1$ represents a sign ambiguity when writing the dilaton exponential term in terms of $\J$. Fortunately, the only %dependence on $\kappa$ is in the overall prefactor, which we will drop from now on. 
For the particular case of $j=0$ the above integral can be computed 
 explicitly,  e.g., by expanding in powers of $\rho$, integrating term by term and then resumming. We obtain
\begin{eqnarray}
&&{\cal C}=-8 \pi \frac{\J_r}{3(1-\J_r^2)^3(b^6 -4 b^2 \rho^2)^2} \Big(\big[(\J_r^4-1 )b^4+2 \J_r^2(7+\J_r^2)\rho^2 \big] E(1-\J_r^{-2})\no \\
 && \ \ \ \ \ \ \ \ \ \ \ \ \ \ \ \ \ \ \ \ \ \ \ \ -2\big[(\J_r^2-1)b^4+(3+5 \J_r^2)\rho^2\big] K(1-\J_r^{-2}) \Big) \ , \label{largea} 
\end{eqnarray}
where we have introduced 
\be 
b^2=1+h^2+\rho^2 \ , \ \ \ \ \ \ \ \ \ \ \ \ 
\J_r= \J \frac{b^2}{\sqrt{b^4-4\rho^2}} \la{kqw}
\ee
 $E$ and $K$ denote the complete elliptic integrals. 
For general $j$, the integral  \rf{kwe} can be computed in some special  limits. For instance, for $\J\to 0 $ we reproduce 
 the well known result for a single circular Wilson loop without angular momentum. 
%. This is of course the case since in this limit there is no motion on the $S^5$.

Another special  case  %which can be analyzed 
is the large $j$ limit in which 
 we can compute the integral by a saddle point approximation.
 For instance, if $h=\rho=0$, then the saddle point is at $u=1/\sqrt{\J}$,  giving
% We obtain a pretty neat result
\begin{equation}
{\cal C}=\frac{4 \J^{1/2}\pi^{3/2}}{(1+\J)^{2+j/2}j^{1/2}}e^{ -1/2   j  K(1-\J^{-2})}\ , 
\end{equation}
where  $K$ is  the complete elliptic integral.
%that came from $\tau_f$ in \rf{tauf}.
 This result is in perfect agreement with the numerics.

\newpage 

%%%%%%%%%%%%%%%%%%%%%%%%%%%%%

\end{document}